\documentclass[12pt]{article}
\usepackage{geometry} % see geometry.pdf on how to lay out the page. There's lots.
\usepackage{amsmath}
\usepackage{graphicx}

\usepackage[comma,authoryear]{natbib}
\begin{document}
\title{The Stark effect in the Bohr-Sommerfeld theory \\ and in Schr\"odinger's  wave mechanics}
%Kramers and Schr\"odinger on the Stark effect in hydrogen

\author{Anthony Duncan and Michel Janssen}

\maketitle

\section{Introduction} 
In the same year that Niels \citet{Bohr 1913} showed that his model of the hydrogen atom correctly reproduces the frequencies of the lines of the Balmer series in the hydrogen spectrum, Johannes \citet{Stark 1913} published his detailed measurements of the splitting of these spectral lines when a hydrogen atom is placed in an external electric field. The Stark effect, as it quickly came to be called, the splitting of spectral lines by electric fields, is the electric analogue of the Zeeman effect, the splitting of spectral lines by magnetic fields, discovered by Pieter Zeeman in 1896 \citep{Kox 1997}. Stark recalled that at a dinner party at Heike Kamerlingh Onnes's house during a visit to Leyden shortly after he discovered the effect, the hostess was seated right  between Zeeman and himself. This prompted a risqu\'e joke on the part of another dinner guest, Paul Ehrenfest, who quipped: ``Well, Mrs.\ Onnes, now you have a choice: do you want to be split electrically or magnetically?" \citep[p.\ 13]{Hermann 1965b}. 

Both splittings won their discoverers a Nobel prize, Zeeman in 1902, Stark in 1919. In the case of the Zeeman effect, Zeeman shared the prize with Hendrik Antoon Lorentz, whose electron theory could account for Zeeman's original findings though not for the more complicated manifestations of the effect found in subsequent years. Three years before Stark won his Nobel prize, Paul \citet{Epstein 1916a, Epstein 1916b} and Karl  \citet{Schwarzschild 1916} showed that Stark's findings could be accounted for on the basis of  Arnold Sommerfeld's (1915a,b) extension of Bohr's theory (Kragh, 2012, pp.\ 154--156; Eckert, 2013a, sec.\ 4.2, pp.\ 44--48). Although the Nobel prize was awarded to Stark alone, part of the significance of the Stark effect was undoubtedly that it supported the Bohr-Sommerfeld theory. Stark, however, was a staunch opponent of the theory \citep[pp.\ 127--128]{Kragh 2012}. He actually spent part of his Nobel lecture railing against it (ibid., pp.\ 168--169).
%\citep[pp.\ 168--169]{Kragh 2012}. 
We will draw the veil of charity over this sad production  and focus instead on Epstein and Schwarzschild.

Not coincidentally, as we will see, Epstein, a Polish-born Russian citizen who had come to Munich in 1910 and taken his doctorate with Sommerfeld in 1914, and Schwarzschild, director of the Astrophysical Observatory in Potsdam, arrived at virtually identical accounts of the Stark effect at almost exactly the same time. Using Sommerfeld's extension of Bohr's theory, especially the notion of (as we would now call it) degeneracy that came with the introduction of multiple quantum numbers, and some powerful techniques from celestial mechanics,  they derived the energy levels for a hydrogen atom in an electric field to first order in the field strength, examined the transitions between these energy levels, and found what most experts, {\it pace} Stark, considered excellent agreement with Stark's spectroscopic data. 

This explanation of the (first-order) Stark effect was hailed, both at the time and by later commentators (see, e.g., Jammer, 1966, pp.\ 108--109; Pais, 1991, p.\ 183), as one of the signature achievements of the old quantum theory of Bohr and Sommerfeld. As Epstein put it in the concluding paragraph of the short note in which he first announced his explanation of the Stark effect:
\begin{quotation}
\noindent
We believe that the reported results prove the correctness of Bohr's atomic model with such striking evidence that even our conservative colleagues cannot deny its cogency. It seems that the potential of the quantum theory in its application to this model is almost miraculous and far from being exhausted  \citep[p.\ 150; translation following Jammer, 1966, p.\ 108]{Epstein 1916a}. 
\end{quotation}
Sommerfeld went even further. By the time he published the first edition of {\it Atombau und Spektrallinien} \citep{Sommerfeld 1919}, which was to become the ``Bible" of the old quantum theory \citep{Eckert 2013b}, Hendrik A.\ (Hans) \citet{Kramers 1919}, Bohr's right-hand man in Copenhagen, had shown in his dissertation that, with the help of Bohr's correspondence principle, the Bohr-Sommerfeld theory could also account for the polarization and, at least qualitatively, the intensities of the various components into which an electric field splits the Balmer lines. Sommerfeld ended the final chapter of his book with a section on the work of Epstein, Schwarzschild, and Kramers on the Stark effect and confidently concluded that ``the theory of the Zeeman effect and especially the theory of the Stark effect belong to the most impressive achievements of our field and form  a beautiful capstone on the edifice of atomic physics" \citep[pp.\ 457--458]{Sommerfeld 1919}. In the next and final paragraph of the book, he suggested that the building of atomic physics was now essentially complete and prophesized that a ``proud new wing" for nuclear physics, built on the same plan as the ``edifice of atomic physics," would soon be added (ibid., p.\ 458).\footnote{For discussion of  Sommerfeld's Munich school in theoretical physics, see \citet{Eckert 1993, Eckert 2013c}, \citet{Seth 2010}, and \citet[Ch.\ 3]{Schweber 2012}.}

Within a few years, it was recognized that Sommerfeld's proclamation of success had been premature. The Zeeman effect turned out to be one of the most thorny problems facing the Bohr-Sommerfeld theory. The theory performed much better in the case of the Stark effect. In hindsight, it is clear that this is mainly because the Stark effect, unlike the Zeeman effect, does not involve spin, at least not in the regime of electric fields used by Stark \citep[p.\ 109]{Jammer 1966}. Yet, as we will show in this paper, the old quantum theory's treatment of the Stark effect also had its share of problems, especially when we compare it, as we plan to do here, to the way the effect is handled in wave mechanics. 

Shortly after the advent of wave mechanics and independently of one another, Erwin \citet{Schroedinger 1926} and \citet{Epstein 1926}, who had meanwhile moved from Munich to Pasadena, applied the new theory to the Stark effect. To first order in the strength of the electric field, wave mechanics gives the same splittings of the energy levels as the old quantum theory. However, whereas the old quantum theory required some ultimately arbitrary selection rules in addition to the basic quantum conditions to restrict the allowed energy levels and the allowed transitions between them to eliminate some pathological orbits and to match the experimental data, the new theory gives the correct energy levels and transitions without any further assumptions. Wave mechanics also predicts the polarizations and intensities of the various components without any appeal to the correspondence principle. The predictions for the intensities differed from those of Kramers and agreed much better with the experimental data. 

Schr\"odinger and Epstein both emphasized these two advantages of their new explanation of the Stark effect. Neither of them, however, commented on another advantage, the solution offered by wave mechanics of a more fundamental problem in the old quantum theory's account of the Stark effect. Both Schwarzschild and Epstein in 1916 and Schr\"odinger and Epstein in 1926 used parabolic coordinates to find the allowed energy levels of a hydrogen atom in an electric field. In the old quantum theory, one would expect that, if the electric field is set to zero, the orbits  in parabolic coordinates reduce to those readily found in polar coordinates for the case without an external electric field. However, even though the energy levels of the orbits are the same in the two coordinate systems, the actual orbits are not. Both \citet[p.\ 507]{Epstein 1916b} and \citet[p.\ 284]{Sommerfeld 1923}
%\citet[pp.\ 349--350]{Sommerfeld 1922} 
dutifully recorded this problem but did not pay any further attention to it.  In wave mechanics, as we will see, the embarrassing non-uniqueness of orbits turns into the completely innocuous non-uniqueness of the basis of eigenfunctions in degenerate systems. The old quantum theory's account of the Stark effect thus illustrates graphically one of the theory's most problematic features, a feature eliminated in the transition to modern quantum mechanics, namely the notion that electrons and other particles have well-defined trajectories. 

\section{The Stark effect in the Bohr-Sommerfeld theory}

Shortly after the discovery of the Zeeman effect, the G\"ottingen theoretical physicist Woldemar Voigt started to look into the theoretical possibility of an electric analogue of the effect \citep{Hermann 1965a}. From 1900 to 1906, Stark worked in G\"ottingen, in the same institute as Voigt. During those years he began a series of experiments to measure the effect of an external electric field on the spectra of (mainly) hydrogen and helium. His efforts finally bore fruit in 1913 in Aachen, where he had been appointed professor at the {\it Technische Hochschule} in 1909 (ibid.). Stark found that  spectral lines emitted by hydrogen and helium split into a number of lines when an electric field is applied. In a series of papers published in 1914, Stark (and his co-authors Georg Wendt and Heinrich Kirschbaum) presented more detailed measurements of the effect in hydrogen, helium, and other elements (see the bibliography of Mehra and Rechenberg, 1982, for detailed references). Like the magnetic field in the case of the Zeeman effect, the electric field typically turned spectral lines into multiplets with more than three components. Voigt's theory, like Lorentz's classical theory for the ``normal" Zeeman effect, could only account for a splitting into three components. Moreover, unlike Lorentz's theory, Voigt's theory gave the wrong values for the frequencies of these components. So did a classical theory by \citet{Schwarzschild 1914} based on an analogy, which he would put to better use two years later, between the perturbation of an electron orbit by an electric field and the perturbation of a planetary orbit by a large but distant mass \citep[p.\ 47]{Eckert 2013a}. Early attempts to account for the Stark effect in hydrogen on the basis of Bohr's new quantum model of the hydrogen atom,  by Emil Warburg and Bohr himself, did not fare any better (Hermann 1965a, pp.\ 15--16; Kragh, 2012, pp.\ 128--129; Eckert 2013a, pp.\ 18--20, pp.\ 26--27). As Armin Hermann (1965a) notes: ``Precondition for a successful treatment [of the Stark effect] was the extension of Bohr's idea by Sommerfeld: the addition of elliptical orbits to the circular orbits of atomic electrons" (p.\ 16).

Bohr's original model did not provide any fundamentally new resources for the analysis of line splittings in electric and magnetic fields. An external field will affect the energy of the allowed orbits. If the change in energy of one orbit is different from that of another, this will also change the frequency of the light emitted in a quantum jump from one to the other. So spectral lines would {\it shift}. But how would we explain that they {\it split}? To answer that question in Bohr's original theory, our only option, it seems,  would be to establish that the effect of an external field on the energy of an orbit depends in just the right ways on the orientation of the orbit with respect to the field. Since the orbits in a gas of atoms will have different orientations with respect to the field, we could then use such dependence to explain the splitting of the spectral lines. It is unclear, however, whether that dependence would give us discrete multiplets or just a blurring of the spectral lines. In fact, \citet[p.\ 450]{Sommerfeld 1915a} expected that this approach could only provide a natural explanation of triplets, as in Lorentz's classical explanation of the Zeeman effect. 

Sommerfeld's extension of Bohr's model, by contrast, suggested a whole new kind of explanation of line splittings. The importance of the generalization to elliptical orbits in this context is that it requires {\it two} quantum numbers whereas the circular orbits of Bohr's original model only require one \citep[p.\ 30]{Eckert 2013b}. Sommerfeld thereby introduced the key notion of degeneracy, to use the modern term. The set of allowed Kepler ellipses correspond to the exact same set of energy values as the original set of allowed circular orbits, but the way in which these energy values and thus the transition frequencies are determined by quantum numbers is  different in the two cases. 

For circular orbits, the allowed energy levels are given by\footnote{Here and in the rest of the paper we use our own modernized notation.}
\begin{equation}
E_n =  - \frac{hR}{n^2},
\label{energy levels Bohr}
\end{equation}
where $h$ is Planck's constant, $R$ is the Rydberg constant, and $n$ is a non-negative integer. The frequency $\nu_{n_i \rightarrow n_f}$ of the radiation emitted when an electron jumps from an initial orbit with quantum number $n_i$ to a final orbit with quantum number $n_f < n_i$ is given by $h  \nu_{n_i \rightarrow n_f} = E_{n_i} - E_{n_f}$. In this way Bohr recovered the empirical formula for the frequencies of the spectral lines in the Balmer series in hydrogen:
\begin{equation}
\nu_{n_i \rightarrow n_f} = R \left( \frac{1}{n_f^2} - \frac{1}{n_i^2} \right).
\label{bohr balmer}
\end{equation}
The most striking success of Bohr's model was that the Rydberg constant could be expressed in terms of more fundamental constants:
\begin{equation}
R = \frac{2  \pi^2 \mu \, e^4}{h^3},
\label{Rydberg}
\end{equation}
where $\mu$ and $-e$ are the (reduced) mass and charge of the electron, respectively.

For elliptical orbits, \citet[p.\ 439]{Sommerfeld 1915a} showed, Eq.\ (\ref{energy levels Bohr})  needs to be replaced by 
\begin{equation}
E_{(n_r, n_\varphi)} = -\frac{hR}{(n_r + n_\varphi)^2},
\label{energy levels Sommerfeld}
\end{equation}
where the radial quantum number $n_r$ is a non-negative integer and the azimuthal quantum number $n_\varphi$ is a positive integer. Eq.\ (\ref{bohr balmer}) consequently needs to be replaced by
\begin{equation}
\nu_{(n_r, n_\varphi)_i \rightarrow (n_r, n_\varphi)_f} = R \left( \frac{1}{(n_r+ n_\varphi)_f^2} - \frac{1}{(n_r + n_\varphi)_i^2} \right),
\label{sommerfeld balmer}
\end{equation}
where $(n_r + n_\varphi)_i > (n_r + n_\varphi)_f$. Sommerfeld found that, unless he quantized eccentricity as well as angular momentum, he did not get a discrete set of energy values for the allowed elliptical orbits. Commenting on Eq.\ (\ref{sommerfeld balmer}), he wrote: 
\begin{quotation}
\noindent
With the addition of our quantized elliptical orbits, the [Balmer] series has gained nothing in terms of number of lines and lost nothing in terms of sharpness. Instead of the diffuse bands discussed earlier [before eccentricity was quantized] we once again have the discrete Balmer lines,  but now with an extraordinarily increased multiplicity of ways in which they can be generated \citep[p.\ 440]{Sommerfeld 1915a}.
\end{quotation}
Sommerfeld only found new lines when he solved the Kepler problem relativistically  in the next paper he presented to the Munich Academy \citep[this paper was presented in January 1916 but was still included in the proceedings volume for 1915]{Sommerfeld 1915b}. The fine structure of the hydrogen spectrum predicted by this relativistic calculation was confirmed within a few months and in close consultation with Sommerfeld by the spectroscopist Friedrich Paschen in his laboratory in T\"ubingen  \citep[pp.\ 49--51]{Eckert 2013a}. Compared to this triumph, the {\it non}-relativistic treatment of the Kepler problem was disappointing:
% as it did not predict any new lines. As Michael Eckert puts it:
\begin{quotation}
\noindent
As long as Sommerfeld could not produce any tangible evidence [i.e., new lines] for the generalized Balmer formula [Eq.\ (\ref{sommerfeld balmer})], his theory compared to Bohr's atomic model had to appear as a very interesting but basically unnecessary extension \citep[p.\ 33]{Eckert 2013a}.
\end{quotation}
As Sommerfeld clearly realized, however, and as Michael Eckert proceeds to show, the extension from circular to elliptic orbits was of great importance even in the absence of ``tangible evidence" deciding between Eq.\ (\ref{bohr balmer}) and Eq.\ (\ref{sommerfeld balmer}). 

Since the various energy levels in a hydrogen atom could be realized in many more ways with Sommerfeld's ellipses than with Bohr's circles, Sommerfeld's theory provided a brand new tool for attempts to account for the Stark and Zeeman effects. The notion of degeneracy, which Sommerfeld in effect introduced by replacing Eq.\ (\ref{energy levels Bohr}) with Eq.\ (\ref{energy levels Sommerfeld}), suggested that one try to explain these effects by showing that electric and magnetic fields lift the degeneracy in the energy of the orbits in just the right way. After all, an aggregate of hydrogen atoms with electrons jumping between all these different allowed elliptic orbits should be expected to start emitting light at many more frequencies than those given by the Balmer series as soon as an electric or a magnetic field changes the energies of those orbits and changes them in a way that is different from one orbit to another so that the radiation frequencies corresponding to transitions between orbits also change. Hence, even in the absence of ``tangible evidence," Sommerfeld's generalization from circular to elliptic orbits had great heuristic potential. 

\citet[pp.\ 449--451]{Sommerfeld 1915a} emphasized this potential in a section of his paper devoted to the Stark effect. Although he acknowledged that a detailed theory of how the electric field lifts the degeneracy in this case had yet to be developed, he pointed to the large number of lines that Stark had found and argued that this made the  approach he was proposing especially promising. ``The hour has come for a true theory of the Zeeman effect," he enthusiastically wrote to Wilhelm Wien on December 31, 1915 (quoted in Eckert, 2013a, p.\ 44), a few weeks after submitting the first and a few weeks before submitting the second paper on his extension of Bohr's theory to the Munich Academy \citep{Sommerfeld 1915a, Sommerfeld 1915b}. Sommerfeld turned out to be wrong about the Zeeman effect, but right about the Stark effect. By the end of March 1916, Epstein and Schwarzschild had worked out a theory of the Stark effect exploiting his notion of degeneracy.

The other key insight that made it possible to account for the Stark effect was Schwarzschild's realization that the quantum conditions proposed by Sommerfeld could be connected to action-angle variables and Hamilton-Jacobi theory, both of which Schwarz\-schild was intimately familiar with because of his expertise in celestial mechanics. 

Drawing on Max Planck's idea of quantizing the phase space spanned by a coordinate $q$ and its associated momentum $p$, \citet[p.\ 429]{Sommerfeld 1915a} had quantized what he called the ``phase integral" for periodic systems, initially for systems with only one degree of freedom,
\begin{equation}
\int p \, dq = nh.
\label{phase integral quantization}
\end{equation}
The integral is to be taken over one period of the motion. The quantum number $n$ has to be a non-negative integer. In this way Sommerfeld could recover, in just a few lines and in a unified way, both the quantization of the energy of the harmonic oscillator needed in black-body radiation theory and the quantization of angular momentum in the Bohr model of the hydrogen atom.

Consider a harmonic oscillator, a point mass with mass $m$ on a spring with spring constant $k$. The characteristic angular frequency of this system is $\omega = 2 \pi \nu = \sqrt{k/m}$. The trajectory of the point mass in the phase space spanned by its position $q$ and its momentum $p$ is an ellipse the size of which is determined by the energy $E = p^2/(2m) + k q^2/2$. Using that $p = \sqrt{2mE}$ for $q=0$ and that $q = \sqrt{2E/k}$ for $p=0$, we see that the major and minor semi-axes, $d_{\rm major}$ and $d_{\rm minor}$, of this ellipse are $\sqrt{2mE}$ and $\sqrt{2E/k}$. The phase integral over one period of the motion is equal to the area of this ellipse, $\pi d_{\rm major} d_{\rm minor}$. Sommerfeld's phase integral quantization condition (\ref{phase integral quantization}) thus gives
\begin{equation}
\int p \, dq = \pi \sqrt{2mE} \sqrt{2E/k} = 2 \pi E \sqrt{m/k} = E/\nu = nh,
\label{Planck}
\end{equation}
which is the familiar quantization condition $E = nh\nu$ for the energy of the harmonic oscillator. 

In the first installment of his famous  trilogy, \citet{Bohr 1913} had quantized the kinetic energy, $E_{\rm kin} = (n/2) h \nu_n$, to select the allowed circular orbits of an electron in a hydrogen atom (with radii $r_n$ and orbital frequencies $\nu_n$). The relation between the kinetic energy of an electron in the $n$th orbit ($n = 1, 2, 3, \ldots$), $E_{\rm kin} = \frac{1}{2}m(2 \pi \nu_n r_n)^2$, and its angular momentum in that orbit, $L=m(2 \pi \nu_n r_n)r_n$, is simply $L = E_{\rm kin}/(\pi \nu_n)$. As \citet[p.\ 15]{Bohr 1913} noted, his quantization condition can thus be written as $L=n \hbar$, where $\hbar \equiv h/2\pi$ \citep[p.\ 280]{Heilbron and Kuhn 1969}. When $(p, q)$ are chosen as  $(L, \varphi)$, Sommerfeld's phase integral quantization condition (\ref{phase integral quantization}) reproduces Bohr's quantization condition in this form:
\begin{equation}
\int_0^{2\pi} L \, d\varphi = 2\pi L = nh.
\end{equation}
Note, however, that we need to add to Sommerfeld's quantum condition (\ref{phase integral quantization}) in this case that $n \neq 0$. There obviously cannot be an orbit with vanishing angular momentum.

Since the Kepler problem involves two degrees of freedom, two phase integrals need to be quantized for the generalization from circular to elliptical orbits \citep[pp.\ 432--440]{Sommerfeld 1915a}. Solving the Kepler problem in polar coordinates, Sommerfeld arrived at the quantum numbers $n_\varphi$ and $n_r$ given in Eq.\ (\ref{sommerfeld balmer}):
\begin{equation}
\int p_\varphi \, d\varphi= n_\varphi h, \quad \int p_r \, dr = n_r h,
\label{Kepler problem in polar coordinates}
\end{equation}
with $p_\varphi \equiv L$ and the additional condition $n_\varphi \neq 0$ (cf.\ Eq.\ (\ref{energy levels Sommerfeld})). \citet{Sommerfeld 1915b} then applied this same approach to the relativistic Kepler problem. As he told Schwarzschild in a letter of December 28, 1915, he was ``moving full steam ahead on spectral lines, with fairy-tale results" \citep[p.\ 29]{Eckert 2013a}. 

Two months later, on March 1, 1916, Schwarzschild sent Sommerfeld a letter in which he made the connection between  phase integrals such as those in Eqs.\ (\ref{phase integral quantization})--(\ref{Kepler problem in polar coordinates}) and action-angle variables \citep[pp.\ 44--45, where this letter is quoted in full]{Eckert 2013a}. Consider Hamilton's equations for some multiply-periodic system with Hamiltonian $H$ described in terms of generalized coordinates $q_k$ and their conjugate momenta $p_k$:
\begin{equation}
 \dot{q}_k = \frac{\partial H}{\partial p_k}, \quad \dot{p}_k = -\frac{\partial H}{\partial q_k}.
\label{Hamilton eqns q p}
\end{equation}
One way to solve these equations is to perform a canonical transformation to new variables, called action-angle variables and typically denoted by $J$ and $w$, that have the desirable property that the Hamiltonian, written as a function of the new variables, only depends on the new momenta, the action variables $J_k$, and not on the new coordinates, the angle variables $w_k$. A generating function $S(q_k, J_k)$, which is known as Hamilton's principal function and turns out to be equal to the action integral for the system, is used to implement the transformation $(q_k, p_k) \longrightarrow (w_k, J_k)$: 
%\citep[p.\ 431]{Goldstein}:
\begin{equation}
w_k = \frac{\partial S}{\partial J_k}, \quad p_k = \frac{\partial S}{\partial q_k}.
\label{generating function}
\end{equation}
In action-angle variables, Hamilton's equations have the simple form:
\begin{equation}
\dot{w}_k = \frac{\partial H}{\partial J_k} = \nu_k, \quad \dot{J}_k = -\frac{\partial H}{\partial w_k} = 0,
\label{Hamilton eqns w J}   
\end{equation}
where the $\nu_k$'s are the characteristic frequencies of the system. The equations for $\dot{w}_k$ allow us to find these characteristic frequencies without fully solving the equations of motion. This explains much of the appeal of action-angle variables in celestial mechanics. Eqs.\ (\ref{Hamilton eqns w J}) can readily be solved.
%are much easier to solve than Eqs.\  (\ref{Hamilton eqns q p}). 
The hard part is finding the generating function $S(q_k, J_k)$ that gets us from Eqs.\ (\ref{Hamilton eqns q p}) to Eqs.\  (\ref{Hamilton eqns w J}). This requires the solution of the so-called Hamilton-Jacobi equation for the system, which we obtain by making the substitutions 
\begin{equation}
p_k  \longrightarrow \frac{\partial S}{\partial q_k}
\label{substitutions}
\end{equation}
(see Eq.\ (\ref{generating function})) in the Hamiltonian $H(q_k,p_k)$ and setting the result equal to some constant.\footnote{For further discussion of canonical transformations, action-angle variables, and Hamilton-Jacobi theory, we refer the reader to graduate textbooks in classical mechanics \citep{Goldstein, Matzner and Shepley 1991, Corben and Stehle}. For an insightful discussion of the use of these techniques in the old quantum theory and wave mechanics, see \citet[Chs.\ 10--11, pp.\ 97--126]{Yourgrau and Mandelstam 1979}.} 

The equations for $\dot{J}_k$ in Eqs.\ (\ref{Hamilton eqns w J}) tell us that the action variables $J_k$ are constants of the motion. This makes them suitable candidates to subject to quantum conditions.  In fact, what Schwarzschild pointed out to Sommerfeld was precisely that his phase integrals can be seen as action variables. Sommerfeld's quantization conditions can be written as:
\begin{equation}
J_k = \int p_k \, dq_k = \int \frac{\partial S}{\partial q_k} dq_k = n_k h.
\label{Sommerfeld-Schwarzschild}
\end{equation}
As Schwarzschild told Sommerfeld, it was only after he had cast the quantization conditions in this new form that they had become truly compelling for him. He added that they now also provided a definite point of departure for the treatment of the Stark effect and the Zeeman effect. ``There are violins hanging all over the quantum heavens," he rhapsodized in another letter to Sommerfeld four days later \citep[p.\ 45]{Eckert 2013a}.

Given how important we now know these techniques from celestial mechanics were for the development of the Bohr-Sommerfeld theory, Sommerfeld's reaction to Schwarz\-schild's communication may come as a surprise. In his reply of March 9, he admitted that he was unfamiliar with the techniques Schwarzschild was referring to and that this would probably be true for most physicists (ibid., p.\ 46). Sommerfeld, however, immediately recognized the importance of  Schwarzschild's intelligence. He relayed the information to Epstein, now an enemy alien in wartime Munich, who, at Sommerfeld's suggestion, had taken up the problem of the Stark effect for a habilitation thesis.  In his interview for the {\it Archive for History of Quantum Physics} (AHQP) in 1963, Epstein recalled the sinking feeling he had upon hearing that Schwarzschild had resumed work on the Stark effect: ``Now I have no prospects unless Schwarzschild should go to Heaven" (session 1, p.\ 11, quoted by Mehra and Rechenberg, 1982, p.\ 225, note 355). Epstein would obviously have preferred Schwarzschild to fiddle with another problem in his quantum heaven, but he may not have known back in March 1916 that Schwarzschild had contracted pemphigus while serving on the Russian front, an auto-immune disease causing painful blisters on the skin that would kill him only two months later. 

Whether or nor he was aware of his rival's predicament, Epstein understood that there was no time to waste if he wanted to beat Schwarzschild to the punch. Fortunately,  inspired perhaps by Schwarzschild's 1914 paper on the Stark effect, Epstein had already begun to bone up on celestial mechanics. On March 21, 1916, he handed in his solution for the Stark effect to Sommerfeld. Later that same day, Sommerfeld received a letter from Schwarzschild with a virtually identical solution \citep[p. 47]{Eckert 2013a}. 

Epstein submitted a preliminary note to {\it Physikalische Zeitschrift} on March 29 \citep{Epstein 1916a}, the day before Schwarzschild submitted his paper to the Berlin Academy \citep{Schwarzschild 1916}. Epstein's note appeared on April 15, Schwarzschild's on May 11 \citep[p.\ 225]{Mehra and Rechenberg 1982}. Schwarzschild died that same day. He was only 42 years old. In a popular article published later that year and entitled ``The quantum theory of spectral lines and the last paper of Karl Schwarzschild," Sommerfeld highlighted and praised Schwarzschild's contributions to the old quantum theory. 

A few days before Schwarzschild had ``gone to heaven," Epstein had submitted a lengthy paper with the details of his explanation of the Stark effect to {\it Annalen der Physik}. This paper appeared in July 1916 \citep{Epstein 1916b}. In what follows, we present the derivation of the formula for the energy levels in the (first-order) Stark effect in the form in which it appears in the dissertation by \citet[pp.\ 16--18]{Kramers 1919}. Kramers (ibid., p.\ 17) cites Epstein's {\it Annalen} paper as his source.
%we follow this paper by Epstein, which, understandably given the circumstances, is superior to Schwarzschild's.

In Cartesian coordinates $(x, y, z)$, the Hamiltonian for an electron in a hydrogen atom in an external electric field ${\mathcal E}$ in the $z$-direction is given by 
\begin{equation}
H = \frac{p^2}{2 \mu} - \frac{e^2}{r} + {\mathcal E}z,
\label{Hamiltonian cartesian}
\end{equation}
where $p^2 \equiv p_x^2 + p_y^2 + p_z^2$, with $(p_x, p_y, p_z)$ the components of the momentum ${\bf p}$, and $r \equiv x^2 + y^2 + z^2$. 
The external electric field applied by Stark was weak compared to that of the hydrogen nucleus keeping the electron in orbit, which means that it can be treated as a small perturbation, amenable to the standard techniques of canonical perturbation theory borrowed from celestial mechanics. 

However, the allowed energy levels of this system could not be found with these techniques in either Cartesian or polar coordinates. Instead, both Epstein and Schwarzschild used parabolic coordinates $(\xi, \eta, \varphi)$, related to $(x, y, z)$ via 
\begin{equation}
z = \frac{\xi - \eta}{2}, \;\;\; x+ iy = \sqrt{\xi \eta} e^{i \varphi}, \;\;\; r = \frac{\xi + \eta}{2},
\label{parabolic coordinates}
\end{equation}
where we followed \citet[p.\ 17, Eq.\ 43]{Kramers 1919} rather than \citet[p.\ 492, Eqs.\ 19--20]{Epstein 1916b}. In parabolic coordinates, the Hamiltonian in Eq.\ (\ref{Hamiltonian cartesian}) is given by:
\begin{equation}
H = \frac{1}{2\mu} \left( \frac{4}{\xi + \eta} ( p_\xi \, \xi \, p_\xi \;) +  \frac{4}{\xi + \eta} (p_\eta \, \eta \, p_\eta) + \frac{1}{\xi \eta} \, p^2_\varphi \right) - \frac{2e^2}{\xi + \eta} +  \frac{1}{2} e{\mathcal E}(\xi - \eta),
\label{Hamiltonian parabolic}
\end{equation}
where $(p_\xi, p_\eta, p_\varphi)$ are the momenta conjugate to $(\xi, \eta, \varphi)$. In the old quantum theory, as in classical mechanics, $p_\xi \, \xi \, p_\xi = \xi \, p_\xi^2$ and $p_\eta \, \eta \, p_\eta = \eta \, p_\eta^2$. It is with malice aforethought that we wrote these products the way we did in Eq.\ (\ref{Hamiltonian parabolic}): in wave mechanics $p_\xi$ is replaced by a differential operator---$(\hbar/i)$ times differentiation with respect to $\xi$---that does not commute with multiplication by $\xi$.

Setting $H = \alpha_1$, where $\alpha_1$ is some negative constant giving the energy of the system, multiplying both sides by $2\mu (\xi + \eta)$, using that
\begin{equation}
\frac{\xi + \eta}{\xi \eta} = \frac{1}{\xi} + \frac{1}{\eta}, \quad (\xi + \eta)(\xi - \eta) = \xi^2 - \eta^2,
\label{xi & eta}
\end{equation}
and making the substitutions
\begin{equation}
p_\xi \longrightarrow \frac{\partial S}{\partial \xi}, \;\;\; p_\eta \longrightarrow \frac{\partial S}{\partial \eta}, \;\;\; p_\varphi \longrightarrow \frac{\partial S}{\partial \varphi},
\label{p to dS/dq}
\end{equation}
we arrive at the Hamilton-Jacobi equation for this system in parabolic coordinates:
\begin{equation}
4 \xi \left( \frac{\partial S}{\partial \xi} \right)^2 + 4 \eta \left( \frac{\partial S}{\partial \eta} \right)^2 + \left( \frac{1}{\xi} + \frac{1}{\eta} \right) \left( \frac{\partial S}{\partial \varphi} \right)^2 - 4\mu e^2 +  \mu e{\mathcal E}(\xi^2 - \eta^2) = 2\mu (\xi + \eta) \alpha_1.
\label{HJ eq}
\end{equation}
At this point, the reason for using parabolic coordinates becomes clear: the Hamilton-Jacobi equation is {\it separable} in these coordinates, which means that its solution is the sum of three terms that each depend only on one of the three coordinates:
\begin{equation}
S(\xi, \eta, \varphi) = S_\xi(\xi) + S_\eta(\eta) + S_\varphi(\varphi).
\label{separation}
\end{equation}
$S_\varphi(\varphi)$ can simply be set equal to $\alpha_3 \varphi$. Hence,
\begin{equation}
\frac{d S_\varphi}{d \varphi} = \alpha_3.
\label{dS/dphi}
\end{equation}
When $\alpha_3$ is substituted for $\partial S/\partial  \varphi =d S_\varphi/d  \varphi $ in Eq.\ (\ref{HJ eq}), the equation splits into a part depending only on $\xi$ and a part depending only on $\eta$. Since the sum of these two parts must vanish, the two parts themselves must be equal but opposite constants. Denoting these constants by $\mp \alpha_2$, we can schematically, write the Hamilton-Jacobi equation (\ref{HJ eq}) as
\begin{equation}
\underbrace{{\rm Terms \; with} \; \frac{d S_\xi}{d \xi}, \alpha_1, \alpha_3  \; {\rm depending \; on \;} \xi}_{\displaystyle = - 2 \alpha_2} + \underbrace{{\rm Terms \; with \;}\frac{d S_\eta}{d \eta}, \alpha_1, \alpha_3 \; {\rm depending \; on \;} \eta }_{\displaystyle = 2 \alpha_2} = 0,
\label{HJ xi eta}
\end{equation}
which splits into separate equations for $S_\xi$ and $S_\eta$ of the form
\begin{equation}
\frac{d S_\xi}{d \xi} = u(\xi, \alpha_1, \alpha_2, \alpha_3), \quad \frac{d S_\eta}{d \eta} = v(\eta,  \alpha_1, \alpha_2, \alpha_3).
\label{dS/dxi dS/deta}
\end{equation}

We now impose the Sommerfeld-Schwarzschild quantum conditions (\ref{Sommerfeld-Schwarzschild}). So far, it may have looked as if we could impose these conditions in arbitrary coordinates. It turns out, however, that the conditions can only be imposed consistently in coordinates in which the Hamilton-Jacobi equation for the system under consideration is separable. As Albert \citet{Einstein 1917} pointed out, this amounts to a severe limitation of the formalism of the Bohr-Sommerfeld theory, over and above its restriction to multiply-periodic systems, as there are many  systems for which the Hamilton-Jacobi equation is not separable in {\it any} coordinates. The formalism, however, does work for the case at hand. Introducing the notation $I_\xi$, $I_\eta$, and $I_\varphi$ for the action variables, we thus impose the quantum conditions
\begin{eqnarray}
I_\xi = \int p_\xi \, d\xi =  \int \frac{d S_\xi}{d \xi} \, d\xi = n_\xi h, \nonumber \\
I_\eta = \int p_\eta \, d\eta = \int \frac{d S_\eta}{d \eta} \, d\eta = n_\eta h, \label{I-xi I-eta I-phi} \\
I_\varphi = \int p_\varphi \, d\varphi = \int \frac{d S_\varphi}{d \varphi} \, d\varphi = n_\varphi h, \nonumber
\end{eqnarray}
where $n_\xi$, $n_\eta$, and $n_\varphi$ are non-negative integers. Both Epstein and Schwarzschild assumed that the values of $I_\xi$, $I_\eta$, and $I_\varphi$ in the presence of a weak electric field ${\mathcal E}$ are the same as their values in the absence of such a field. Where the cases ${\mathcal E}=0$ and ${\mathcal E} \neq 0$ differ is in how the separation constants depend on the action variables. One of these separation constants, $\alpha_1$, is equal to the energy $E$. So even though the action variables have the same values for ${\mathcal E}=0$ and ${\mathcal E} \neq 0$, the energy  does not. 

The justification of the assumption that action variables have the same values for ${\mathcal E}=0$ and ${\mathcal E} \neq 0$ is that they are what Ehrenfest called adiabatic invariants. In June 1916, \citet{Ehrenfest 1916a} presented a paper to the Amsterdam academy connecting the adiabatic principle, which he had already been working on for a number of years, to the Bohr-Sommerfeld theory \citep{Perez 2009}. In July, he submitted a similar paper to {\it Annalen der Physik} \citep[see P\'erez, 2009, pp.\ 83--84, for a concise overview of Ehrenfest's papers on the topic in 1916]{Ehrenfest 1916b}. In September he added a postscript to this paper responding to Schwarzschild's combination of the Bohr-Sommerfeld theory and Hamilton-Jacobi theory:
\begin{quotation}
\noindent
The beautiful researches of Epstein, Schwarzschild, and others [such as Peter Debye \citep[p.\ 52]{Eckert 2013a}] which have appeared in the meantime, show the great importance that cases integrable by means of St\"ackel's [1891] method of ``separation of the variables" have for the development of the theory of quanta. Hence the question arises: to what extent are the different parts into which these authors separate the integral of action according to St\"ackel's method adiabatic invariants? \citep[translation based on P\'eres, 2009, p.\ 93]{Ehrenfest 1916b}.
\end{quotation}
Ehrenfest's question was taken up by one of his students in Leyden, Johannes (Jan) \citet{Burgers 1917a, Burgers 1917b, Burgers 1917c}, who showed that action variables such as those in Eqs.\ (\ref{I-xi I-eta I-phi}) are indeed adiabatic invariants (Klein 1970, pp.\ 290--291; Yourgrau and Mandelstam 1979, pp.\ 110-111; P\'eres 2009, pp.\ 93--102).  

We now return to the calculation for the Stark effect. The next step is to evaluate the integrals in Eqs.\ (\ref{I-xi I-eta I-phi}) after substitution of the right-hand sides of Eqs.\ (\ref{dS/dphi}) and (\ref{dS/dxi dS/deta}) for the integrands. For $I_\varphi$, we find with the help of Eq.\ (\ref{dS/dphi}):
\begin{equation}
I_\varphi  = \int \frac{d S_\varphi}{d \varphi} \, d\varphi = 2 \pi \alpha_3 = n_\varphi h.
\label{I-phi}
\end{equation}
In other words, $\alpha_3 = n_\varphi \hbar$, so $n_\varphi$ is the familiar azimuthal quantum number typically denoted nowadays by $m$. Similarly, although performing the integrals now requires some effort, we can express the action variables $I_\xi$ and $I_\eta$ in terms of the separation constants $\alpha_1$, $\alpha_2$, and $\alpha_3$. We then invert these relations to find the $\alpha$'s in terms of the $I$'s and thereby in terms of the  quantum numbers $n_\xi$, $n_\eta$, and $n_\varphi$. We need to do this twice, first for ${\mathcal E}=0$, then to first order in ${\mathcal E} \neq 0$. We will not go through these calculations in detail; we will only state the end results.

In the absence of an external field (${\mathcal E}=0$), the sum of $I_\xi$, $I_\eta$, and $I_\varphi$ for ${\mathcal E}=0$ is given by
\begin{equation}
I_\xi + I_\eta + I_\varphi  = \frac{2 \pi \mu e^2}{\sqrt{- 2 \mu \alpha_1}}.
\label{sum I-xi I-eta I-phi 0}
\end{equation}
Solving for $\alpha_1$, we find
\begin{equation}
\alpha_1 = - \frac{2 \pi^2 \mu e^4}{(I_\xi + I_\eta + I_\varphi)^2}.
\label{alpha 1 no E}
\end{equation}
This reduces to the expression $-hR/n^2$ for the allowed energy levels in a hydrogen atom in the absence of an electric field found earlier (see Eqs. (\ref{energy levels Bohr}) and (\ref{energy levels Sommerfeld})), if the sum of the quantum numbers  introduced in Eqs.\ (\ref{I-xi I-eta I-phi}) is set equal to the principal quantum number $n$:
\begin{equation}
n = n_\xi + n_\eta + n_\varphi.
\label{n in old theory}
\end{equation}
As in Sommerfeld's calculation for elliptical orbits in polar coordinates (cf.\ Eq.\ (\ref{energy levels Sommerfeld})), the calculation for elliptical orbits in parabolic coordinates for ${\mathcal E} = 0$ thus leads to the same energy levels as Bohr's original calculation for circular orbits (cf.\ Eq.\ (\ref{energy levels Bohr})) but does reveal the degeneracy of those energy levels:
\begin{equation}
E_{(n_\xi, n_\eta, n_\varphi)} = \alpha_1 = - \frac{2 \pi^2 \mu e^4}{h^2 (n_\xi + n_\eta + n_\varphi)^2} = -\frac{hR}{n^2},
\label{energy levels Epstein no E}
\end{equation}
where in the last step we used expression (\ref{Rydberg}) for the Rydberg constant. As in Sommerfeld's formula for the allowed energy levels in polar coordinates (see Eq.\ (\ref{energy levels Sommerfeld})), we need to impose  further restrictions on the allowed values of  the quantum numbers (\ref{I-xi I-eta I-phi}) \citep[sec.\ 4, pp.\ 497--501]{Epstein 1916b}. First, $n_\xi$, $n_\eta$, and $n_\varphi$ cannot all three be zero as the principal quantum number $n$ would then be zero. Second, even if $n_\xi \neq 0$ and/or $n_\eta \neq 0$, $n_\varphi$ cannot be zero. As long as ${\mathcal E}=0$ there is no problem, but when ${\mathcal E} \neq 0$ this orbit becomes unstable and the electron will eventually hit the nucleus. 

The degeneracy in the energy levels in Eq.\ (\ref{energy levels Epstein no E}) is lifted once the electric field is switched on.  The integrals in Eqs.\ (\ref{I-xi I-eta I-phi}) now have to be evaluated to first order in ${\mathcal E}$ (where in terms of order ${\mathcal E}$ we can use the relations between $\alpha$'s and $I$'s found for ${\mathcal E} = 0$). In this approximation, Eq.\ (\ref{energy levels Epstein no E}) gets replaced by
\begin{equation}
E_{(n_\xi, n_\eta, n_\varphi)} = \alpha_1 =  -\frac{hR}{n^2} +  C {\mathcal E} n ( n_\xi - n_\eta),
\label{energy levels Epstein E}
\end{equation}
where $n = n_\xi + n_\eta + n_\varphi$ and $C \equiv 3 h^2/8 \pi^2  e \mu$. The electric field thus produces the splittings 
\begin{equation}
\Delta E_{(n_\xi, n_\eta, n_\varphi)} = C {\mathcal E} n ( n_\xi - n_\eta)
\label{delta E}
\end{equation}
of the energy levels \citep[p.\ 508, Eq.\ 62]{Epstein 1916b} and the splittings 
\begin{equation}
\Delta \nu_{(n_\xi, n_\eta, n_\varphi)_i \longrightarrow (n_\xi, n_\eta, n_\varphi)_f} = 
\frac{C {\mathcal E}}{h} \, \Big( [n ( n_\xi - n_\eta)]_i - [n ( n_\xi - n_\eta)]_f \Big)
\label{delta nu}
\end{equation}
of the transition frequencies (ibid., p.\ 509, Eq.\ 65). The splittings $\Delta E$ in Eq.\ (\ref{delta E}) are symmetric around the values for $E$ without an external field. The splittings $\Delta \nu$ in Eq.\ (\ref{delta nu}) are likewise symmetric around the values for $\nu$ without an external field. This is in accordance, as \citet[pp.\ 509--510]{Epstein 1916b} noted, with Stark's experimental results.

\begin{figure}[h]
   \centering
   \includegraphics[width=5in]{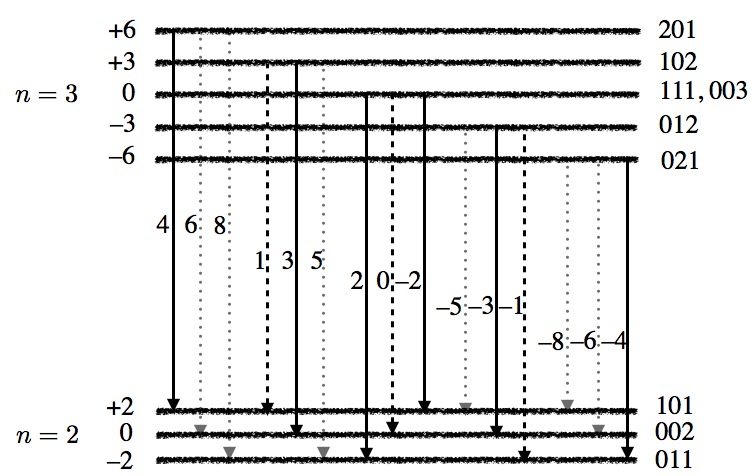} 
   \caption{{\small Stark effect for the Balmer line $H_\alpha$ in the hydrogen spectrum: splittings $\Delta E$ (in units of $C {\mathcal E}$ with ${\mathcal E}$ the strength of the electric field and $C \equiv 3 h^2/8 \pi^2  e \mu$) of energy levels for $n=2$ and $n=3$ [horizontal lines with values of $\Delta E$ to the left and values of quantum numbers $(n_\xi, n_\eta, n_\varphi)$ to the right]; splittings $\Delta \nu$ (in units of $C {\mathcal E}/h$) of the frequency of the radiation emitted in  transitions from $n=3$ to $n=2$ [arrows with values of $\Delta \nu$ to the left---solid arrows: parallel polarization; dashed arrows: perpendicular polarization; dotted arrows: violation of  selection rule]. The figure is not drawn to scale: the energy gap between the $n=2$ and $n=3$ levels is much greater than the level splittings.}}
   \label{splittings}
\end{figure}

Fig.\ \ref{splittings} (based on the numbers in Epstein, 1916b, p.\ 512, Table I) illustrates the Stark effect for the Balmer line $H_\alpha$ in the hydrogen spectrum. It shows the splittings $\Delta E$ of the energy levels  $n=2$ and $n=3$ in the presence of an external electric field of strength ${\mathcal E}$ and  the splittings $\Delta \nu$ of the frequencies of the radiation emitted in transitions from $n=3$ to $n=2$. Similar though increasingly more complicated diagrams can be drawn for  $H_\beta$ ($n=4 \longrightarrow n=2$),  $H_\gamma$ ($n=5 \longrightarrow n=2$), and $H_\delta$ ($n=6 \longrightarrow n=2$) \citep[pp.\ 512--513, Tables II--IV]{Epstein 1916b}.

The electric field splits the lower  level ($n=2$) into three levels. For  ${\mathcal E} = 0$, the energies of the orbits picked out by the values $(101)$, $(002)$, and $(011)$  for the quantum numbers $(n_\xi, n_\eta, n_\varphi)$ are all the same. For ${\mathcal E} \neq 0$, the energy of the orbit  $(101)$ is raised by $2C {\mathcal E}$, while the energy of the orbit $(011)$ is lowered by that same amount (cf.\ Eq.\ (\ref{delta E})). The electric field splits the upper  level ($n=3$) into five levels. For  ${\mathcal E} = 0$, the energies of the orbits $(201)$, $(102)$, $(111)$, $(003)$, $(012)$, and $(021)$ are all the same. For ${\mathcal E} \neq 0$, the energies of the orbits $(102)$ and $(201)$ are raised by $3C {\mathcal E}$ and $6C {\mathcal E}$, respectively, while the energies of the orbits $(012)$ and $(021)$ are lowered by those same amounts. 

For ${\mathcal E} = 0$, a quantum jump of an electron from any of the six possible $n=3$ orbits to any of the three possible $n=2$ orbits is accompanied by the same energy loss and therefore by emission of radiation of the same frequency. For ${\mathcal E} \neq 0$, as indicated by the arrows in Fig.\ \ref{splittings}, the energy loss in a quantum jump from $n=3$ to $n=2$ can take on fifteen different values, resulting in frequency shifts $\Delta \nu$ ranging from $-8C {\mathcal E}/h$ to $+8C {\mathcal E}/h$. This means that the frequency of the Balmer line $H_\alpha$, emitted in the transition from $n=3$ to $n=2$, splits into fifteen different frequencies. Illustrating Epstein's general observation noted above, the fourteen shifted frequencies lie symmetrically on opposite sides of the unshifted one. 

Epstein eliminated six of these fourteen shifted frequencies, three on each side of the unshifted one. He adopted, at least initially, a selection rule proposed by \citet[pp.\ 447--448]{Sommerfeld 1915a}, which requires that
\begin{equation}
n_\xi^f \leq n_\xi^i, \quad n_\eta^f \leq n_\eta^i, \quad n_\varphi^f \leq n_\varphi^i
\label{Sommerfeld selection rule}
\end{equation}
\citep[p.\ 511, Eq.\ (69)]{Epstein 1916b}. According to this selection rule, the six transitions in which one of the three quantum numbers increases are forbidden. These are the transitions represented by dotted arrows in Fig.\ \ref{splittings}.\footnote{Under this selection rule, the transitions `$(003) \! \longrightarrow \! (011)$' and `$(003) \! \longrightarrow \! (101)$' are also forbidden but $E_{(003)} = E_{(111)}$ even if ${\mathcal E} \neq 0$ and the transitions `$(111) \! \longrightarrow \! (011)$' and `$(111) \! \longrightarrow \! (101)$' are allowed, so this does not affect the number of lines (cf. Fig.\ \ref{splittings}).}  The corresponding lines were either absent or exceedingly faint in Stark's spectroscopic data, which supported Sommerfeld's selection rule. The nine remaining transitions all matched lines clearly present in Stark's data: the six transitions  indicated by solid arrows (with $\Delta \nu$ equal to $\pm 2$, $\pm 3$, $\pm 4$ times $C {\mathcal E}/h$) producing light polarized parallel to the field; the three transitions  indicated by dashed arrows (with $\Delta \nu$ equal to $0$, $\pm 1$ times $C {\mathcal E}/h$) producing light polarized perpendicular to the field.

The splittings of other Balmer lines found by Stark violated Sommerfeld's selection rule. To match Stark's data, \citet[p.\ 516]{Epstein 1916b} eventually settled on a modified version of the rule, 
\begin{equation}
n_\xi^f \leq n_\xi^i, \quad n_\eta^f \leq n_\eta^i, \quad n_\varphi^f \leq n_\varphi^i + 1,
\label{Epstein selection rule}
\end{equation}
and emphasized that transitions violating this rule are not strictly forbidden, just highly improbable. As we mentioned in the introduction, the explanation of the Stark effect in the Bohr-Sommerfeld theory thus requires what in the final analysis are rather arbitrary restrictions, both on the allowed energy levels (see our comments following Eq.\ (\ref{energy levels Epstein no E})) and on the allowed transitions between them (see Eqs.\ (\ref{Sommerfeld selection rule}) and (\ref{Epstein selection rule})).

With the help of these additional conditions, Schwarzschild and Epstein could account for the frequencies of all the components into which the Balmer lines were observed to split in the Stark effect. This was rightfully celebrated as a tremendous success for the Bohr-Sommerfeld theory. However, neither Schwarzschild nor Epstein could account for the polarizations or the intensities of these components.  \citet[sec.\ 8, pp.\ 514--518]{Epstein 1916b} devoted a section of his paper to
 polarizations and intensities. He began that section with the following disclaimer:
\begin{quotation}
\noindent
The theory of Bohr's atomic model in its current form is based on the consideration of stationary orbits at the beginning and at the end of every individual radiation process. What happens during the transition of an electron from one orbit to another is still very unclear to us. Accordingly, the goal of this section is not to draw theoretical conclusions about polarization and intensities \ldots but only to extract lawlike regularities from the available empirical material  \citep[p.\ 514]{Epstein 1916b}. 
\end{quotation}
For the polarizations Epstein stated the following empirical law. Even values of $\Delta n_\varphi \equiv n_\varphi^i - n_\varphi^f$ give rise to parallel polarization, odd values to perpendicular polarization (ibid.). Note that for the dashed arrows in Fig.\ \ref{splittings} (polarization perpendicular to the field),  $\Delta n_\varphi = \pm 1$, while for the solid arrows (polarization parallel to the field), $\Delta n_\varphi = 0$. Turning to intensities, Epstein wrote:
\begin{quotation}
\noindent
[T]he following hypothesis seems to fit the facts best: {\it a component \ldots is stronger, caeteris paribus, the greater the largest of the three differences} [between initial and final quantum numbers] \ldots the idea behind this is that the situation is similar to when there is a difference in altitude: the greater the difference in quantum numbers the greater the tendency to eliminate that difference \citep[p.\ 515, emphasis in the original]{Epstein 1916b}.
\end{quotation}

Bohr's correspondence principle \citep[Ch.\ 5]{Kragh 2012} provided  a much more promising starting point for  dealing with polarizations and intensities of spectral lines than Epstein's ``evening out differences in altitude" analogy. Consider the Fourier expansion of the position ${\bf x}$ of a particle \citep[p.\ 31, Eq.\ (31)]{Bohr 1918}:
\begin{equation}
{\bf x} = \sum_{\tau_1 \ldots \tau_s} {\bf C}_{\tau_1 \ldots \tau_s} \, {\rm cos} \, 2 \pi \Big\{ \big( \tau_1 \omega_1 \ldots \tau_s \omega_s \big) t + c_{\tau_1 \ldots \tau_s} \Big\}.
\label{tau's Bohr}
\end{equation}
Commenting on this expression, Bohr wrote:
\begin{quotation}
\noindent
Now on ordinary electrodynamics the coefficients $C_{\tau_1 \ldots \tau_s}$ in the expression [Eq.\ (\ref{tau's Bohr})] for the displacement of the particles in the different directions would in the well known way determine the intensity and polarization of the emitted radiation of the corresponding frequency $\tau_1 \omega_1 + \ldots \tau_s \omega_s$. As for systems of one degree of freedom we must therefore conclude that, in the limit of large values for the $n$'s, the probability of spontaneous transition between two stationary states of a conditionally periodic system, as well as the polarization of the accompanying radiation, can be determined directly from the values of the coefficient ${\bf C}_{\tau_1 \ldots \tau_s}$ in (31) corresponding to a set of $\tau$'s given by $\tau_k = n_k' - n_k''$, if $n_1', \ldots n_s'$ and  $n_1'', \ldots n_s''$ are the numbers which characterize the two stationary states \citep[pp.\ 31--32]{Bohr 1918}.
\end{quotation}
In other words, Bohr suggested that, in the limit of high quantum numbers, the intensity of the radiation of frequency $\nu_{i \rightarrow f}$ emitted in the transition from an initial orbit with the values $(n_1, n_2, n_3)_i$ for the quantum numbers  to a final orbit with the values $(n_1, n_2, n_3)_f$ should be equal to the square of the coefficient ${\bf C}_{\tau_1 \ldots \tau_s}$ of a term in the Fourier expansion of that orbit such that 
\begin{equation}
\nu_{i \rightarrow f} = \frac{1}{h} \big(E_i - E_f \big) = \sum_{k=1}^s \omega_k \tau_k,
\label{Fourier frequencies}
\end{equation}
and $\tau_k = n^i_k - n^f_k$ (ibid., p.\ 30, Eq.\ 30). The core of Bohr's correspondence principle was to take the leap of faith that this asymptotic connection between his own theory and classical electrodynamics would continue to hold if we go from high to low quantum numbers \citep[and, building on their paper, Bokulich, 2008, sec.\ 4.2, pp.\ 80--94]{Fedak and Prentis 2002}. With this general prescription, both intensities and polarizations could be handled. A transition between two orbits will be accompanied by radiation with a certain polarization and a certain intensity whenever the relevant coefficient in the orbit's Fourier expansion is non-vanishing. For very high quantum numbers it does not matter whether we consider the Fourier expansion of the initial or of the final orbit. For low quantum numbers, however, this does matter, rendering Bohr's prescription ambiguous. Should we consider the Fourier expansion of the initial or of the final orbit? Some average of the two perhaps? Or an average  over initial and final orbit and all orbits in between?

Despite this ambiguity, this approach based on the correspondence principle was much more promising than the one taken by Epstein based on Sommerfeld's selection rule (\ref{Sommerfeld selection rule}), which Bohr rejected:
\begin{quotation}
\noindent
Thus, from the fact that in general negative as well as positive values for the $\tau$'s appear in [Eq.\ (\ref{tau's Bohr})], it follows that we must expect that in general not only such transitions will be possible in which all the $n$'s [e.g., the quantum numbers $n_\xi$, $n_\eta$, and $n_\varphi$] decrease, but that also transitions will be possible for which some of the $n$'s increase while others decrease. This conclusion, which is supported by observations on the fine structure of the hydrogen lines  as well as on the Stark effect, is contrary to the suggestion by Sommerfeld \ldots that every of the $n$'s must remain constant or decrease under a transition
\citep[p.\ 32]{Bohr 1918}.
\end{quotation}

In his dissertation, Bohr's assistant \citet{Kramers 1919} adopted and elaborated Bohr's correspondence-principle approach to account for the polarizations and intensities of the various components into which the Balmer lines split in the Stark effect. By distinguishing Fourier expansions of the motion in the direction of the field ($z$) and in the plane perpendicular to the field ($x + i y$), Kramers could account for the polarizations found by Stark. The exponent for the Fourier expansion of $z$ (p.\ 21, Eq.\ (60)), both for initial and final orbit, does not contain $\tau_3$, which according to Bohr's correspondence principle should be set equal to $n^i_\varphi - n^f_\varphi$. This suggests that $\Delta n_\varphi = 0$ for all transitions in which radiation polarized parallel to the field is emitted. Similarly, only terms with $\tau_3=\pm 1$ are present in the exponent of the Fourier expansion of $x + i y$ (p.\ 23, Eq.\ (67)). This suggests that $\Delta n_\varphi = \pm 1$ for all transitions in which radiation polarized perpendicular to the field is emitted. Fig.\ \ref{splittings} illustrates that these conclusions based on the correspondence principle are supported by Stark's findings. The solid-arrow transitions (parallel polarization) all have $\Delta n_\varphi = 0$; the dashed-arrow transitions (perpendicular polarization) all have $\Delta n_\varphi = \pm 1$.

\citet[Tables I--IV on pp.\ 55--57]{Kramers 1919} could also account, at least qualitatively, for the intensities of the various components Stark had found. In principle, Kramers used the average of the squares of coefficients of the relevant Fourier components of the initial and the final orbits to estimate the intensity of the corresponding line. However, even in cases where a certain frequency was completely absent from the Fourier expansion of both the initial and the final orbit, Kramers left open the possibility that the corresponding line might appear in the spectrum, albeit only faintly, as its frequency might be present in the Fourier expansion of some orbit in between. Kramers thus allowed several lines that are forbidden by the selection rules of Sommerfeld and Epstein (see Eqs.\ (\ref{Sommerfeld selection rule}) and (\ref{Epstein selection rule})). As we saw above, Epstein had ruled out six possibilities for the transition $n = 3 \longrightarrow n=2$ (see the dotted arrows in Fig.\ \ref{splittings} with values $\pm 5$, $\pm 6$, $\pm 8$ times $C {\mathcal E}/h$ for $\Delta \nu$). Kramers predicted (correctly as it turned out) that these components of the Stark splitting of $H_\alpha$ had just escaped notice so far because of their low intensity (see Kramers, 1919, Appendix, Fig.\ 1 [Kramers, 1956, p.\ 105])

To conclude our discussion of the Stark effect in the old theory, we turn to the problem mentioned in the introduction that  the electron orbits allowed by the quantization conditions depend on the coordinates in which these conditions are imposed.  Although one finds the same energy levels in different coordinate systems, one does not always find the same orbits \citep[p.\ 101]{Jammer 1966}. The analysis of the Stark effect in hydrogen provides a dramatic illustration of this problem. The orbits found in parabolic coordinates when the electric field is set equal to zero  differ sharply from those found without an external electric field in polar coordinates. Both \citet[p.\ 507]{Epstein 1916b} and \citet[p.\ 284]{Sommerfeld 1923} acknowledged this discrepancy. Both of them expressed the (idle) hope that the problem would disappear once relativistic effects were taken into account. Epstein wrote:
\begin{quotation}
\noindent
Even though this does not lead to any shifts in the line series, the notion that a preferred direction introduced by an external field, no matter how small,  should drastically ({\it in einschneidender Weise}) alter the form and orientation of stationary orbits seems to me to be unacceptable. The solution of this apparent paradox is to be expected from a theory in which relativity and external field are taken into account at the same time \ldots This would involve an extension of the quantum conditions for situations with a superposition of two effects that individually can be handled through a separation of variables \citep[p.\ 507]{Epstein 1916b}.
\end{quotation}
It is not clear what such an extension would look like. Epstein and Sommerfeld may have held out hope that an exact treatment of a system would lead to a Hamilton-Jacobi equation that is only separable in one unique set of coordinates. One could then argue that the real orbits of the system are the ones found in those coordinates. Alas, the exact treatment of any but the simplest systems will result in Hamilton-Jacobi equations that are not separable in {\it any} coordinates.

\begin{table}[h]
\centering
\label{tab:scales}
\begin{tabular}{|c|c|c|c|c|c|}
\hline
\multicolumn{1}{|c|}{$n$}
&\multicolumn{1}{c|}{ $n_{r}$}
&\multicolumn{1}{c|}{ $l$}
&\multicolumn{1}{c|}{ $\epsilon$} \\[2pt]
\hline
1  & 0  & 1 & 0   \\
2  & 0 & 2 & 0  \\
2  & 1 & 1 & $\sqrt{3}/2$ \\
3  & 0 & 3 & 0  \\
3  & 1 & 2 & $\sqrt{5}/3$ \\
3  & 2 & 1 & $2\sqrt{2}/3$  \\
\hline
\end{tabular}
\vspace{.1in}
\caption{{\small Angular momentum ($l$ times $\hbar$) and eccentricity ($\epsilon$) for low-lying orbits in polar coordinates}}
\end{table}

\begin{table}[h]
\centering
\vspace{.1in}
\label{tab:scales}
\begin{tabular}{|c|c|c|c|c|c|}
\hline
\multicolumn{1}{|c|}{$n$}
&\multicolumn{1}{|c|}{$n_\xi$}
&\multicolumn{1}{|c|}{$n_\eta$}
&\multicolumn{1}{c|}{ $n_\varphi$}
&\multicolumn{1}{c|}{$l$}
&\multicolumn{1}{c|}{ $\epsilon$ } \\[2pt]
\hline
1  & 0  & 0 & 1 & 1 & 0   \\
2  & 0 & 0  & 2 & 2 & 0 \\
2  & 1 & 0 & 1  & $\sqrt{2}$ & $1/\sqrt{2}$   \\
2  &  0 & 1 & 1 & $\sqrt{2}$ & $1/\sqrt{2}$ \\
2 & 1 & 1 & 0  &  $2\sin{(\pi\delta)}$ & $\cos{(\pi\delta)}$ \\
3  & 0 & 0 & 3  & 3 & 0 \\
3  & 1 & 1 & 1  & $\sqrt{1+8\sin^{2}{(\pi\delta)}}$ & $2\sqrt{2}\cos{(\pi\delta)}/3$ \\
3  & 1 & 0 & 2  & $\sqrt{6}$  & $1/\sqrt{3}$ \\
3  & 2 & 0 & 1  & $\sqrt{3}$  & $\sqrt{2/3}$ \\
3  & 0 & 1 & 2  & $\sqrt{6}$ & $1/\sqrt{3}$ \\
3  & 0 & 2 & 1  & $\sqrt{3}$ & $\sqrt{2/3}$ \\
\hline
\end{tabular}
\caption{{\small Angular momentum ($l$ times $\hbar$) and eccentricity ($\epsilon$) for low-lying orbits in parabolic coordinates}}
\end{table}

Tables 1 and 2 illustrate just how strongly the orbits in parabolic coordinates differ from those in polar coordinates. The first three columns in Table 1 and the first four columns in Table 4 give the quantum numbers---$(n_r, l)$ and $(n_\xi, n_\eta, n_\varphi)$, respectively---for orbits with principal quantum number $n =1, 2, 3$ (where $n = n_r + l = n_\xi + n_\eta + n_\varphi$). The last two columns in both tables give the values for the angular momentum and the eccentricity for the orbits characterized by these quantum numbers. These entries are based on the following relations, which we will just state here rather than derive. The size $l$ of the angular momentum in units of $\hbar$ and the eccentricity $\epsilon$ are related via
\begin{equation} 
l = n\sqrt{1-\epsilon^{2}}, \quad \epsilon = \sqrt{ 1-\frac{l^{2}}{n^{2}}}. 
\label{l, n, and e}
\end{equation}
We used the latter expression to find the numbers in the column for $\epsilon$ in Table 1. To find the corresponding entries in parabolic coordinates, we introduce the quantities $\sigma_1$ and $\sigma_2$:
\begin{equation}
\sigma_{1} \equiv \frac{1}{n}\sqrt{n_\xi(n_\xi+n_\varphi)}, \quad  \sigma_{2} \equiv \frac{1}{n}\sqrt{n_\eta(n_\eta+n_\varphi)}.
\label{sigma}
\end{equation}
The eccentricity can be written as a function of these quantities and a phase parameter $\delta$, which can take on a continuum of values:
\begin{equation}
\epsilon = \sqrt{\sigma_{1}^{2}+\sigma_{2}^{2}+2\sigma_{1}\sigma_{2}\cos{(2\pi\delta)}}.
\label{e parabolic}
\end{equation}
We used this expression to find the numbers in the column for $\epsilon$ in Table 2. To obtain the values in the column for $l$, we used the expression obtained by substituting Eq.\ (\ref{e parabolic}) for $\epsilon$ into the first of Eqs.\ (\ref{l, n, and e}). 

Comparing Tables 1 and 2, we see that only the circular orbits (with $n_r = n_\xi = n_\eta = 0$) are the same in polar and parabolic coordinates. All other orbits are different. The appearance of the phase $\delta$ in Table 2 shows that, in many cases, the values of $(n_\xi, n_\eta, n_\varphi)$ do not even pick out discrete orbits but rather continuous sets of orbits. Orbits were abandoned in the transition from the old quantum theory to matrix mechanics in 1925, largely because of problems in dispersion theory \citep{Duncan and Janssen 2007}. In hindsight, we can see that one of the most celebrated successes of the old quantum theory, the Stark effect, should have made proponents of the theory suspicious of the notion of well-defined electron orbits in atoms well before 1920. 

\section{The Stark effect in Schr\"odinger's wave mechanics}

The derivation of the formula for the Stark effect in wave mechanics shows a strong family resemblance to the derivation of \citet{Epstein 1916b} and \citet{Schwarzschild 1916} in the old quantum theory. Independently of one another, \citet{Schroedinger 1926} and \citet{Epstein 1926} applied the new wave mechanics to the Stark effect. Schr\"odinger's paper was published in {\it Annalen der Physik} on July 13. Epstein's paper is signed July 29 and was published in {\it Physical Review} in October 1926. Epstein had moved to Pasadena in 1921. In his paper, \citet[p.\ 695, note 1]{Epstein 1926} cited Schr\"odinger's first and second ``communication'' ({\it Mitteilung}) on wave mechanics as well as Schr\"odinger's paper on the equivalence of wave and matrix mechanics (which appeared in May 1926), but not the third communication. Presumably, the July 13 issue of the {\it Annalen} had not reached Pasadena by July 29.

In the abstract of his paper, Epstein emphasized the advantages of the new theory of the Stark effect over the old one:\begin{quotation}
\noindent
(1) {\it Positions of lines} practically coincide with those obtained in the writer's old theory which gave an excellent agreement with experiment. (2) {\it Intensity expressions} are obtained in a simple closed form: (a) Components which, in the old theory, had to be ruled out by a special postulate now drop out automatically. (b) The computed intensities of the remaining components check the observed within the limits of experimental error \citep[p.\ 695]{Epstein 1926}. 
\end{quotation}
In the introduction, he elaborated:
\begin{quotation}
\noindent
The positions of the lines turn out to be practically the same as in the writer's old theory. The first order terms are identical with the old expressions, the second order terms [which we are ignoring in this paper (AD \& MJ)] show a very slight difference. The main interest of the paper lies, therefore, in the intensity formulas, which are remarkably simple in their structure and agree with the observed values better than Kramers' intensity expressions derived from Bohr's correspondence principle (ibid.).
\end{quotation}

To bring out the close analogy between the calculations in the old and the new quantum theory, we sketch the derivation of the formula for the Stark effect in hydrogen in wave mechanics (see, e.g., Condon and Shortley, 1963, pp.\ 398--404, for a modern textbook treatment that follows Schr\"odinger and Epstein). As in the old quantum theory, the starting point is the Hamiltonian (\ref{Hamiltonian parabolic}) in parabolic coordinates. Instead of the substitutions (\ref{p to dS/dq}) of $\partial S/\partial \xi$ for $p_\xi$ etc., we now make the substitutions
\begin{equation}
p_\xi \longrightarrow \frac{\hbar}{i} \frac{\partial}{\partial \xi}, \quad p_\eta \longrightarrow \frac{\hbar}{i} \frac{\partial}{\partial \eta}, \quad p_\varphi \longrightarrow \frac{\hbar}{i} \frac{\partial}{\partial \varphi},
\label{p to operator p}
\end{equation}
to form the Hamilton operator $\hat{H}$ entering into the time-independent Schr\"odinger equation,
\begin{equation}
\hat{H} \psi = \alpha_1 \psi,
\label{Schroedinger eq}
\end{equation}
where $\psi(\xi, \eta, \varphi)$ is the wave function in parabolic coordinates. Following the notation used in our calculation in the old quantum theory, we use $\alpha_1$ to label the energy eigenvalues. With the substitutions (\ref{p to operator p}) the Hamiltonian (\ref{Hamiltonian parabolic}) becomes the Hamilton operator
\begin{equation}
\hat{H} = - \frac{\hbar^2}{2\mu} \left( \frac{4}{\xi + \eta} \left( \frac{\partial}{\partial \xi} \xi \frac{\partial}{\partial \xi} \right) +  \frac{4}{\xi + \eta} \left( \frac{\partial}{\partial \eta} \eta \frac{\partial}{\partial \eta} \right) + \frac{1}{\xi \eta} \frac{\partial^2}{\partial \varphi^2}  \right)
- \frac{2e^2}{\xi + \eta} +  \frac{1}{2} e{\mathcal E}(\xi - \eta).
\label{Hamilton operator}
\end{equation}
Inserting this expression into Eq.\ (\ref{Schroedinger eq}), dividing both sides by $\psi$ and multiplying by $2\mu (\xi + \eta)$ (using relations (\ref{xi & eta})), we arrive at:
\begin{equation}
- \frac{\hbar^2}{\psi} \left( 4 \frac{\partial}{\partial \xi} \xi \frac{\partial}{\partial \xi} +  4   \frac{\partial}{\partial \eta} \eta \frac{\partial}{\partial \eta} +   \left( \frac{1}{\xi} + \frac{1}{\eta}  \right) \frac{\partial^2}{\partial \varphi^2} \right) \psi - 4 \mu e^2  +  \mu e {\mathcal E}(\xi^2 - \eta^2)  =  2\mu (\xi + \eta) \alpha_1.
\label{Schroedinger eq H hydrogen}
\end{equation}
Note the similarity between the Schr\"odinger equation (\ref{Schroedinger eq H hydrogen}) and the Hamilton-Jacobi equation (\ref{HJ eq}) in the old quantum theory. Hamilton-Jacobi theory played an important role in the development of wave mechanics. It was the embodiment of the optical-mechanical analogy that guided Schr\"odinger's search for a new wave mechanics underlying ordinary particle mechanics the way wave optics underlies ray optics \citep{Joas and Lehner 2009}. Schr\"odinger's account of the Stark effect shows that the connection between wave mechanics and Hamilton-Jacobi theory also enabled him to transfer important mathematical techniques from the old quantum theory to his new theory.

Eq.\ (\ref{Schroedinger eq H hydrogen}), like Eq.\ (\ref{HJ eq}), is separable in parabolic coordinates. In the case of the Schr\"odinger equation, this means that its solution has the form
\begin{equation}
\psi(\xi, \eta, \varphi) = \psi_\xi(\xi) \psi_\eta(\eta) \psi_\varphi(\varphi).
\label{factorization}
\end{equation}
The wave function $\psi$ and the generating function $S$ are related via $\psi = e^{{\displaystyle iS/\hbar}}$ (as noted, for instance, in the opening paragraph of Pascual Jordan's (1927) {\it Neue Begr\"undung} paper discussed in Duncan and Janssen, 2013). Hence, if $S$ is the sum of three functions, each of which depends on only one of the three coordinates $\xi$, $\eta$, and $\varphi$ (see Eq.\ (\ref{separation})), $\psi$ will be the product of three such functions:
\begin{equation}
\psi(\xi, \eta, \varphi) = e^{i(S_\xi(\xi) + S_\eta(\eta) + S_\varphi(\varphi))/\hbar} = \psi_\xi(\xi) \psi_\eta(\eta) \psi_\varphi(\varphi),
\label{psi & S}
\end{equation}
Just as we could set $S_\varphi(\varphi)= \alpha_3 \varphi$ (see Eq.\ (\ref{I-phi})), we can set $\psi_\varphi(\varphi) = e^{i\alpha_3 \varphi/\hbar}$, with $\alpha_3 = n_\varphi \hbar$ and $n_\varphi = m$. Upon substitution of $-m^2 \psi$ for $\partial^2 \psi/\partial \varphi^2 = d^2 \psi_\varphi/d\varphi^2$ in Eq.\ (\ref{Schroedinger eq H hydrogen}), we are left with an equation that splits into a part depending only on $\xi$ and a part depending only on $\eta$. Both parts must therefore be constant. Denoting these constants by $\mp 2 \alpha_2$ as we did in the corresponding Eq.\ (\ref{HJ xi eta}) in the old quantum theory, we arrive at equations of the form
\begin{equation}
\xi  \frac{d^2\psi_\xi}{d \xi^2} + \frac{d\psi_\xi}{d \xi} +  \big( \ldots \big) \psi_\xi = 0, \quad \eta \frac{d^2\psi_\eta}{d \eta^2} + \frac{d\psi_\eta}{d \eta} +  \big( \ldots \big) \psi_\eta =0.
\label{psi xi eta eq}
\end{equation}
The expressions in parentheses are functions of $\xi$ and $\eta$, respectively, containing the separation constants $\alpha_1$, $\alpha_2$, $\alpha_3$, and the field strength ${\mathcal E}$. 

As in the old quantum theory, we first solve these equations for ${\mathcal E} =0$ and then to first (and second) order in ${\mathcal E}$. For our purposes, the first step, with ${\mathcal E} =0$, turns out to be the most interesting one, and we will focus on that part of the calculation. We begin by studying the  behavior of $\psi_\xi(\xi)$ and $\psi_\eta(\eta)$ at small and large $\xi$ and $\eta$, respectively. This leads us to write these functions in the form
\begin{equation}
\psi_\xi(\xi) = \xi^{|m|/2} e^{- \xi/2na} f(\xi), \quad \psi_\eta(\eta) = \eta^{|m|/2} e^{- \eta/2na} g(\eta),
\label{psi xi psi eta factorized}
\end{equation}
where $f$ and $g$ are as yet unknown functions. In the process we replaced $\alpha_1$ by 
\begin{equation}
na \equiv \frac{\hbar}{\sqrt{-2 \mu \alpha_1}} \;\;\; {\rm with} \; a \equiv \frac{\hbar^2}{\mu e^2},
\label{n & a}
\end{equation}
anticipating that $n$ will eventually be identified as the principal quantum number. Inserting Eqs.\ (\ref{psi xi psi eta factorized}) for $\psi_\xi$ and $\psi_\eta$ into Eqs.\ (\ref{psi xi eta eq}), we find equations for $f$ and $g$ of the form:
\begin{equation}
\xi f'' + \big( \ldots \big) f' + \big( \ldots \big) f = 0, \quad \eta g'' + \big( \ldots \big) g' + \big( \ldots \big) g = 0.
\label{eq for f}
\end{equation}
The solution of these equations will be polynomials in $\xi$ and $\eta$, respectively:
\begin{equation}
f(\xi) = \sum_{k =0}^{n_\xi} a_k \, \xi^k, \quad g(\eta)= \sum_{l=0}^{n_\eta} b_l \, \eta^l,
\label{polynomials}
\end{equation}
with recursion relations on their coefficients (of the form $a_{k+1}/a_k = \ldots$ and $b_{l+1}/b_l = \ldots$). For the wave function to be square-integrable, the polynomials in Eq.\ (\ref{polynomials}) have to break off at some point, i.e., there must be values $n_\xi$ and $n_\eta$ of $k$ and $l$ such that $c_{n_\xi+1}\! = \! 0$ and $c_{n_\eta+1} \!= \! 0$. This leads to the conditions:
\begin{equation}
n_\xi = \frac{n}{2} \left(1 \! - \!  \frac{|m| \! + \! 1}{n} \right) - \frac{\alpha_2na}{2 \hbar^2}, \quad
n_\eta = \frac{n}{2} \left(1 \! - \!  \frac{|m| \! + \! 1}{n} \right) + \frac{\alpha_2na}{2 \hbar^2}.
\label{quantum conditions}
\end{equation}
Combining these two conditions, we find
\begin{equation}
\frac{n}{2} \left(1 \! - \!  \frac{|m| \! + \! 1}{n} \right) - n_\xi = n_\eta - \frac{n}{2} \left(1 \! - \!  \frac{|m| \! + \! 1}{n} \right),
\label{n xi n eta 1}
\end{equation}
or, equivalently,
\begin{equation}
n = n_\xi + n_\eta + |m| + 1.
\label{n in new theory}
\end{equation}
Comparing this result in wave mechanics with the corresponding result (\ref{n in old theory}) in the old quantum theory, we notice that the difference between the two results is the final term $+1$ in Eq.\ (\ref{n in new theory}). This extra term obviates the need for a special condition to rule out $|m| =0$. Both \citet[p.\ 463]{Schroedinger 1926} and Epstein emphasized this point. Elaborating on point (2a) in his abstract (quoted above), Epstein commented:
\begin{quotation}
\noindent
It will be remembered that the restriction for the azimuthal quantum number [$|m| > 0$] was an additional one, not following from the dynamical conditions. It was introduced by Bohr for the purpose of eliminating plane orbits, moving in which the electrons would sooner or later undergo a collusion [sic] with the nucleus. In our new theory an additional restriction is not necessary \citep[p.\ 708]{Epstein 1926}.
\end{quotation}

In the so-called WKB(J) approximation---developed independently, shortly after the formulation of wave mechanics, by Gregor \citet{Wentzel 1926}, L\'eon \citet{Brioullin 1926}, and \citet{Kramers 1926}, and earlier, in a different context, by Harold \citet{Jeffreys 1924}---conditions similar to the quantum conditions (\ref{Sommerfeld-Schwarzschild}) of the Bohr-Sommerfeld theory emerge from the requirement that different parts of the approximate solutions of the Schr\"odinger equation constructed according to the WKB(J) method merge properly. In the regime of high quantum numbers, one  finds conditions of the form $\int p_i \, dq_i = (n_i + \alpha) h$ in this way, where $\alpha$ is equal to $\frac{1}{4}$ times an integer (now called the Maslov index;  Gutzwiller, 1990, p.\ 211).
 %\citep[p.\ 211]{Gutzwiller 1990}. 
In the analysis of the Stark effect in parabolic coordinates, $\alpha$ turns out to be equal to $\frac{1}{2}$ and the quantum numbers $n_\xi$ and $n_\eta$ in the old quantum theory are replaced by $n_\xi + \frac{1}{2}$ and $n_\eta + \frac{1}{2}$, respectively. This explains the extra term 1 in Eq.\ (\ref{n in new theory}) for the principal quantum number.

Wave mechanics gives much better results than the old quantum theory for the intensities of the various lines in the Stark effect. \citet[pp.\ 709--710]{Epstein 1926} included some tables in his paper comparing the intensities predicted by wave mechanics to those observed. ``We see that the agreement is fair," he concluded, ``and decidedly better than that obtained from Bohr's correspondence principle in Kramers' work" (ibid., p.\ 710).  Schr\"odinger reached the same conclusion. Despite the notorious animosity between proponents of wave and matrix mechanics, Schr\"odinger, who had just published his paper on the equivalence of the two formalisms, borrowed freely from matrix mechanics to calculate intensities. As he explained at the beginning of the section on intensities in his paper:
\begin{quotation}
\noindent
According to Heisenberg, if $q$ is a Cartesian coordinate, the square of the matrix element \ldots should be a measure for the ``transition probability from the $r^{\rm th}$ to the $r'^{\rm th} state$," more precisely speaking the intensity of that part of the radiation connected to this transition that is polarized in the direction of $q$ \citep[p. 465]{Schroedinger 1926}.
\end{quotation}
In modern Dirac notation, this matrix element would be written as 
\begin{equation}
\langle n^{r'}_\xi, n^{r'}_\eta, m^{r'} | \, q \, | n^{r'}_\xi, n^{r}_\eta, m^{r} \rangle
\label{matrix element}
\end{equation}

Neither Epstein nor Schr\"odinger seems to have realized that the new account of the Stark effect was superior to the old one in yet another respect. As we mentioned  in the introduction, quantum mechanics replaces the embarrassing non-uniqueness of orbits in the old quantum theory (see Tables 1 and 2 at the end of Section 2) by a completely innocuous non-uniqueness of bases of eigenfunctions. Consider, for example, the three orbits for the lower level ($n=2$) in Fig.\ \ref{splittings}, with the values $(011)$, $(002)$, and $(101)$ for the quantum numbers $(n_\xi, n_\eta, n_\varphi)$ associated with the use of parabolic coordinates in the old quantum theory. In wave mechanics, these three levels do not correspond to orbits but to three (orthonormal) wave functions characterized by those same quantum numbers. All three are eigenfunctions (at least to first order in the electric field strength ${\mathcal E}$) of the Hamiltonian (\ref{Hamilton operator}) for the hydrogen atom in an external electric field. These wave functions can be written as linear combinations of three (orthonormal) wave functions characterized by quantum numbers in polar coordinates. As long as there is  no external electric field (${\mathcal E} =0$), these wave functions are eigenfunctions of the Hamiltonian as well. However, as soon as the  field is switched on (${\mathcal E} \neq 0$), they no longer are; only linear combinations of these wave functions in polar coordinates that correspond to the eigenfunctions of the Hamiltonian in parabolic coordinates are. This is no problem at all. In the old quantum theory, we get a different set of allowed physical states (represented by orbits in a miniature solar system) depending on whether we use polar or parabolic coordinates. In the new quantum theory, we get the same set of allowed physical states (now represented by wave functions or, more generally, by vectors or rays in Hilbert space) regardless of which coordinates we use. We just have the freedom of writing any state as a linear combination of any orthonormal basis of states in the Hilbert space.

\section{Conclusion: Stark contrasts between the old and the new quantum theory}

The explanation of the Stark effect by Epstein and Schwarzschild in 1916 was a triumph for the old quantum theory. In {\it Atombau und Spektrallinien}, from which we already quoted a few passages in the introduction, Sommerfeld wrote that this explanation was in such complete agreement with the empirical data that ``any doubt about the correctness and uniqueness of the solution is no longer possible" \citep[p.\ 440]{Sommerfeld 1919}. ``[T]he classical theory," he pointed out, ``failed completely in the case of the Stark effect. By contrast, the quantum theory fully reproduces the observations in all their rich detail (including recently the polarizations)"  (ibid.). A few pages later, after covering the work of Epstein, Schwarzschild, and Kramers on the Stark effect, he wrote in the concluding paragraphs of his book:
\begin{quotation}
\noindent
The frequencies, especially those of the electric splittings could be derived with extraordinary certainty and completeness from the principles of Bohr's theory of the hydrogen atom and of quantum emission. With a sensible extension of the theory, [the polarization] could also be explained in a way that hardly leaves any gaps. The ravine that originally seemed to open up between the quantum theory and the wave theory of spectral lines could therefore on essential points already be bridged. Not much is missing for it to be definitively filled in. In this sense, the theory of the Zeeman effect and especially that of the Stark effect belong to the most impressive achievements of our field and form a beautiful capstone on the edifice of atomic physics \citep[pp.\ 457--458]{Sommerfeld 1919}.
\end{quotation}
As we showed in section 2, however, even the Stark effect revealed some serious cracks in Sommerfeld's edifice. To account for the effect in the old quantum theory, one had to make some arbitrary assumptions in addition to the Bohr-Sommerfeld quantum conditions to rule out certain orbits. To calculate intensities of lines on the basis of Bohr's correspondence principle, one had to make at least one more arbitrary assumption. It was not enough to stipulate that the intensity of a line of a given frequency is given by the square of the coefficient of the term in the Fourier expansion of the orbit with that frequency. One also had to decide whether to use the Fourier expansion of the initial orbit, the final orbit, or some weighted average of both and everything in between. Moreover, contrary to the calculations of the frequencies of the various lines, the calculations of their intensities only gave limited agreement with the (admittedly also less secure) experimental data.  Most worrisome of all, we saw that the actual orbits predicted by the old quantum theory depend on the coordinates chosen to impose the quantum conditions (see Tables 1 and 2). Both Sommerfeld and Epstein clearly identified this problem but their response to it was little more than wishful thinking that the problem would somehow go away.

As we showed in section 3, all these problems were solved in 1926 when Schr\"odinger and Epstein explained the Stark effect on the basis of the new wave mechanics. The old explanation was certainly helpful as the mathematical techniques needed to solve the problem in the two theories are very similar, as we also saw in section 3. In particular, it suggested that the Schr\"odinger equation, like the Hamilton-Jacobi equation, would be separable in parabolic coordinates (cf.\ Eqs.\ (\ref{HJ eq})--(\ref{dS/dxi dS/deta}) and Eqs.\ (\ref{Schroedinger eq H hydrogen})--(\ref{psi xi eta eq})). The new theory determines all allowed states and transitions without any additional assumptions. In particular, the principal quantum number picked up an extra term of +1 (see Eq.\ (\ref{n in new theory})), which obviated the need to rule out certain combinations of values of the three parabolic quantum numbers.  Wave mechanics also replaced the ambiguous guidelines based on the correspondence principle for calculating intensities by the straightforward and definite prescription that intensities are given by the squares of the matrix elements of position, leading to results that agreed much better with the experimental data. Finally, the embarrassing non-uniqueness of orbits in the old quantum theory was replaced by a completely innocuous non-uniqueness of bases of eigenfunctions in wave mechanics.

The Stark effect is remembered to this day as one of the few admittedly qualified successes of the old quantum theory. We suspect that this is largely because after 1926 it became just one of many unqualified successes of the new quantum theory.

\section*{Acknowledgments}

This paper was written as part of a joint project in the history of quantum physics of the {\it Max-Planck-Institut f\"ur Wissenschaftsgeschichte} and the {Fritz-Haber-Institut} in Berlin. We are grateful to Michael Eckert for helpful discussion of Sommerfeld and to Martin J\"ahnert, Enric P\'erez Canals, Blai Pie Valls, and Robert ``Ryno" Rynasiewicz for helpful discussion of the correspondence principle.

\end{document}